\documentclass[12pt]{article}

\def\mytitle#1{\setcounter{equation}{0}
\setcounter{footnote}{0}
\begin{flushleft}\Large\textbf{#1}\end{flushleft}
\vspace{0.27cm}}
\def\myname#1{\leftline{{\large #1}}\vspace{-0.13cm}}
\def\myplace#1#2{\small\begin{flushleft}\textit{#1}\\
\texttt{#2}\end{flushleft}}

\def\myclassification#1{\small\noindent
Pacs no : 04.20-q, 04.30-w, 04.40.Nr
       #1\vspace{0.5cm}}
\usepackage{graphicx}% Include figure files
\begin{document}

\mytitle{A study of charged cylindrical Gravitational collapse with dissipative fluid}

\vskip0.2cm \myname{Sanjukta Chakraborty\footnote{sanjuktachakraborty77 @gmail.com}}
\vskip0.2cm \myname{Subenoy Chakraborty\footnote{schakraborty@math.jdvu.ac.in}}

\myplace{Department of Mathematics, Tarakeswar Degree College, Tarakeswar, India.}{}
\myplace{Department of Mathematics, Jadavpur University, Kolkata-700 032, India.}{}

%%%%%%%%%%%%%%%%%%%%%%%%%%%%%%%%%%%%%%%%%%%%%%%%%%%%%%%%%%%%%%%%%%%%%%%%%%%%%%%%%%%%%%%%%%%%%%%%%%%%%%%%%%%%%%%%%%%%\begin{abstract}
\begin{abstract}

The present works deals with gravitational collapse of cylindrical viscous heat conducting anisotropic fluid following the work of Misner and Sharp. Using Darmois matching conditions, the dynamical equations are derived and the effect of charge and dissipative quantities over the cylindrical collapse are analyzed. Finally, using the Miller-Israel-Steward causal thermodynamic theory, the transport equation for heat flux are derived and its influence on collapsing system has been studied.  \\
%%%%%%%%%%%%%%%%%%%%%%%%%%%%%%%%%%%%%%%%%%%%%%%%%%%%%%%%%%%%%%%%%%%%%%%%%%%%%%%%%%%%%%%%%%%%%

Keywords : Cylindrical collapse, Dissipation, heat flux, Junction conditions, Dynamical equations .
\end{abstract}
%%%%%%%%%%%%%%%%%%%%%%%%%%%%%%%%%%%%%%%%%%%%%%%%%%%%%%%%%%%%%%%%%%%%%%%%%%%%%%%%%%%%%%%%%%%%
\myclassification{}\\
%%%%%%%%%%%%%%%%%%%%%%%%%%%%%%%%%%%%%%%%%%%%%%%%%%%%%%%%%%%%%%%%%%%%%%%%%%%%%%%%%%%%%%%%%%%%
\section{Introduction}
%%%%%%%%%%%%%%%%%%%%%%%%%%%%%%%%%%%%%%%%%%%%%%%%%%%%%%%%%%%%%%%%%%%%%%%%%%%%%%%%%%%%%%%%%%%%

A challenging but curious issue in gravitational physics as well as in relativistic astrophysics is to know the final fate of a continual gravitational collapse. The stable configuration of a massive star persists as long as the inward pull of gravity is neutralized by the outward pressure of the nuclear fuel at the core of the star. Subsequently, when the star has exhausted its nuclear fuel there is no longer any thermonuclear burning and there will be endless gravitational collapse. However, depending on the mass of the collapsing star, the compact objects such as white dwarfs, neutron stars and black holes are formed. In white dwarf and neutron star gravity is counter balanced by electron and neutron degeneracy pressure respectively while black hole is an example of the end state of collapse.\\

The study of gravitational collapse was initiated long back in 1939 by Oppenheimer and Snyder [1]. They have studied the collapse of a homogeneous spherical dust cloud in the frame work of general relativity. Then after a quarter century, a more realistic investigation was done by Misner and Sharp[2] with perfect fluid in the interior of a collapsing star. In both the studies, the exterior of the collapsing star was chosen as vacuum. Vaidya[3] formulated the non-vacuum exterior of a star having radiating fluid in the interior. An inhomogeneous spherically symmetric dust cloud was analytically studied by Joshi and Singh [4] and they have shown that the final fate of the collapsing star depend crucially on the initial density profile and the radius of the star. Debnath etal[5] investigated collapse dynamics of the non-adiabatic fluid, considering quasi-spherical Szekeres space-time in the interior and plane symmetric Vaidya solution in the exterior region.\\

Although most of the works on collapse dynamics are related to spherical objects, still there are interesting information about self-gravitating fluids for collapsing object with different symmetries. The natural choice for non-spherical symmetry is axis symmetric objects. The vacuum solution for Einstein field equations in cylindrically symmetric space-time was obtained first by Levi-Civita [6] but  still it is a challenging issue of interpreting two independent parameters in the solution. Herrera etal [7] studied cylindrical collapse of non-dissipative fluid with exterior Einstein -Rosen space-time and showed wrongly a non-vanishing radial pressure on the boundary surface and subsequently in collaboration with M.A.H. Maccallum [8] they corrected the  result. Then Herrera and collaborators investigated cylindrical collapse of matter with[9] or without shear [10]. \\

Further, the junction conditions due to Darmois [11] has a very active role in dealing collapsing problems. Sharif etal [12-14] showed the effect of positive cosmological constant on the collapsing process by using junction conditions between static exterior and non-static interior with a cosmological constant. Also Herrera etal [15], using junction conditions were able to prove that any conformally flat cylindrically symmetric static source cannot be matched to the Levi-Civita space-time. Then  Kurita and Nakao [16] formulated naked singularity  along the axis of symmetry, considering cylindrical collapse with null dust.\\

Moreover from realistic point of view it is desirable to consider dissipative matter in the context of collapse dynamics [17-19]. Considering collapse of a radiating star with dissipation in the form of radial heat flow and shear viscosity, Chan[20] has showed that shear viscosity plays a significant role in the collapsing process. Collapse dynamics with dissipation of energy as heat flow and radiation has been studied by Herrera and Santos[18]. Subsequently, by Considering of causal transport equations related to different dissipative components (heat flow, radiation, shear and bulk viscosity) Herrera etal [15,21,22] investigated the collapse dynamics. The same collapsing process with plane symmetric geometry or others  has been examined by Sharif etal [23,24] .\\

On the other-hand, in the context of gravitational waves, the sources must have non spherical symmetry. Further, cylindrical collapse of non-dissipative fluid with exterior containing gravitational waves shows non-vanishing pressure on the boundary surface by using  Darmois matching conditions. Recently, it has been verified [25] in studying cylindrical collapse of anisotropic dissipative fluid with formation of gravitational waves outside the collapsing matter.\\

In the present work, following Misner and Sharp collapse dynamics of viscous, heat conducting charged anisotropic fluid in cylindrically symmetric background will be studied. The paper is organized as follows. Section 2 deals with basic equations related to interior and exterior space-time. The junction conditions are evaluated and discussed in Section 3. The dynamical equations are derived and studied in Section 4. Finally, the process of mass, heat and momentum transfer through transport equation is discussed in section 5. \\

\section{Interior and exterior space-time: Basic equations.}

Mathematically, the whole four dimensional space-time manifold having a cylindrical collapsing process can be written as $M=M^+U \Sigma U M^-$ with $M^{-}\bigcap M^{+}=\phi$. Here, $\Sigma$, the collapsing cylindrical surface is a time-like three surface and is the boundary of the two four dimensional sub-manifolds   $M^-$ (interior) and $M^+$ (exterior).\\
In $M^-$ choosing co-moving coordinates the line element can be written as [25] 
\begin{equation}
d{s_-^2}=-{A^2}d{t^2} +{B^2 }d{r^2} +{C^2}d{\phi^2} +{D^2}d{z^2}
\end{equation}\\
where the metric coefficients are functions of t and r i.e. A=A(t,r) and so on.  
Also due to cylindrical symmetry, the coordinates are restricted as:\\
$-\infty\leq t\leq +\infty,~~~r\geq 0,~~~-\infty<z<+\infty,~~~0\leq \phi \leq 2\pi$\\

For compact notation we write $\lbrace x^{-\mu} \rbrace \equiv[t,r,\phi,z]~~~,~~~ (\mu=0,1,2,3) $.\\

The anisotropic fluid having dissipation in the form of shear viscosity and heat flow has the energy- momentum tensor of the form [7,9]
\begin{equation}
T_{\mu\nu}=(\rho+p_t){v_\mu}{v_\nu}+{p_t}g_{\mu\nu}+({p_r}-{p_t}){\chi_\mu}{\chi_\nu}-2\eta\sigma_{\mu\nu}+2q_{(\mu}v_{\nu)}
\end{equation}\\
Here $\rho,~p_r,~p_t~ ~~\eta~~~ and~~q_{\mu}$ stands for energy density ,the radial pressure, the tangential pressure, coefficient of shear viscosity and radial heat flux vector respectively. Also
 $v_\mu$ and $\chi_\mu$ are  unit time-like and space-like vectors satisfying the following relations
\begin{equation}
v_\mu v^\mu =-\chi_\mu\chi^\mu=-1~~~,~~~~\chi^\mu v_\mu=0~~~,~~~q_{\mu}v^{\mu}=0~~~~
\end{equation}\\
Moreover,  the shear tensor $\sigma_{\mu\nu}$ has the expression 
\begin{equation}
\sigma_{\mu\nu}= v_{(\mu;\nu)}+a_{(\mu}v_{\nu)}-\frac{1}{3}\Theta(g_{\mu\nu}+v_\mu v_\nu)
\end{equation}\\
where $a_\mu=v_{\mu;\nu}v^\nu $ is the acceleration vector and $\Theta=v^\mu;_\mu$ is the expansion scalar. \\
 For the above metric one may choose the unit time-like vector, space-like vector and heat flux vector in a simple form as  
\begin{equation}
v^\mu=A^{-1}\delta_0^\mu~~~,~~~~~~~~~~~~~~\chi^\mu=B^{-1}\delta_1^\mu~~~,~~~~~q^{\mu}=q\delta^{\mu}_{1}~~~~~
\end{equation}\\
The shear tensor has only non-zero diagonal components as   
\begin{equation}
\sigma_{11}=\frac{B^2}{3A}[\Sigma_1-\Sigma_3]~~,~~~~~~\sigma_{22}=\frac{C^2}{3A}[\Sigma_2-\Sigma_1]~~~and~~~\sigma_{33}=\frac{D^2}{3A}[\Sigma_3-\Sigma_2]~~~with~~~~~\sigma^2=\frac{1}{6A^2}[\Sigma_1^2+\Sigma_2^2+\Sigma_3^2]
\end{equation}\\
where
$ \Sigma_1=\frac{\dot{B}}{B}-\frac{\dot{C}}{C}~~,~~~\Sigma_2=\frac{\dot{C}}{C}-\frac{\dot{D}}{D}~~,~~~\Sigma_3=\frac{\dot{D}}{D}-\frac{\dot{B}}{B}
$\\

Also the acceleration vector and the expansion scalar have the explicit expressions
\begin{equation}
a_1=\frac{A^\prime}{A}~~,\Theta=\frac{1}{A}(\frac{\dot{B}}{B}+\frac{\dot{C}}{C}+\frac{\dot{D}}{D})
\end{equation}\\
In the above, by notation we have used
$\cdot$ $\equiv\frac{\partial}{\partial t}$~~~~~and~~~~~$^\prime$ $\equiv\frac{\partial}{\partial r}.$\\
If in addition we assume the above fluid distribution to be charged then the energy-momentum tensor for the electromagnetic field has the form
\begin{equation}
E_{\alpha\beta}=\frac{1}{4\pi}(F^{\alpha}_{\mu}F_{\nu\alpha}-\frac{1}{4}F^{\alpha\beta}F_{\alpha\beta}g_{\mu\nu})~~~~~
\end{equation}

where the Maxwell field tensor $F_{\alpha\beta}$ is related to the four potential $\phi_{\alpha}$ as 
\begin{equation}
F_{\alpha\beta}=\phi_{\beta , \alpha}-\phi_{\alpha , \beta}
\end{equation}
and the evolution of the field tensor corresponds to Maxwell equations
\begin{equation}
F^{\alpha\beta}_{;\beta}=4\pi J^{\alpha}
\end{equation}
where $J^{\alpha}$ the four current vector.\\
As the charge per unit length of the cylinder is at rest with respect to comoving co-ordinates so the magnetic field will be zero in this local coordinate system [26,27]. Hence the four potential and the four current takes the simple form
\begin{equation}
\phi_{\alpha}=\phi\delta^{0}_{\alpha} ,~~~~J^{\alpha}=\epsilon v^{\alpha}
\end{equation}
where $\phi=\phi(t,r)$ is the scalar potential and $\epsilon=\epsilon(t,r)$ is the charge density.\\

From the law of conservation of charge : $J^{\alpha}_{;\alpha}=0$, one obtains the total charge distribution interior to radius $r$ and per unit length of the cylinder as
\begin{equation}
s(r)=2\pi\int^{r}_{0}\epsilon BCD dr
\end{equation}
Now the explicit form of the Maxwell's equations (10) for the interior space-time $M^{-}$ are given by
\begin {equation}
\phi^{''}-(\frac{A^{'}}{A}+\frac{B^{'}}{B}-\frac{C^{'}}{C}-\frac{D^{'}}{D})\phi^{'}=4\pi\epsilon AB^{2}
\end{equation}
and 
\begin{equation}
\dot{\phi}^{'}-(\frac{\dot{A}}{A}+\frac{\dot{B}}{B}-\frac{\dot{C}}{C}-\frac{\dot{D}}{D})\phi^{'}=0
\end{equation}
A first integral of equation (13) gives
\begin{equation}
\phi^{'}=\frac{2sAB}{CD}
\end{equation}
which satisfies identically the other Maxwell's equation(14).
Hence one obtains the electric field intensity as 
$E(t,r)=\frac{s(r)}{2\pi C}$

Further, in the interior space-time $M^{-}$ the Einstein field equations $G_{\alpha\beta}=8\pi(T_{\alpha\beta}+E_{\alpha\beta})$ have the explicit form

\begin{equation}
\frac{A^{2}}{B^{2}}(-\frac{C^{''}}{C}-\frac{D^{''}}{D}+\frac{B^{'}}{B}(\frac{C^{'}}{C}+\frac{D^{'}}{D})-\frac{C^{'}D^{'}}{CD})+(\frac{\dot{B}\dot{C}}{BC}+\frac{\dot{B}\dot{D}}{BD}+\frac{\dot{C}\dot{D}  }{CD})=8\pi(\rho A^{2}-2\eta\sigma_{00})+4\frac{s^{2}A^{2}}{C^{2}D^{2}}
\end{equation}\\

\begin{equation}
-\frac{B^{2}}{A^{2}}(\frac{\ddot{C}}{C}+\frac{\ddot{D}}{D}+\frac{\dot{C}\dot{D}}{CD}-\frac{\dot{A}\dot{C}}{AC}-\frac{\dot{A}\dot{D}}{AD})+(\frac{C^{'}D^{'}}{CD}+\frac{A^{'}C^{'}}{AC}+\frac{A^{'}D^{'}}{AD})=8\pi(p_{r}B^{2}-2\eta\sigma_{11})-4\frac{s^{2}B^{2}}{C^{2}D^{2}}
\end{equation}

\begin{equation}
-\frac{C^2}{A^2}[\frac{\ddot{B}}{B}+\frac{\ddot{D}}{D}-\frac{\dot{A}}{A}(\frac{\dot{B}}{B}+\frac{\dot{D}}{D})+\frac{\dot{B}\dot{D}}{BD}]+\frac{C^2}{B^{2}}[\frac{A^{''}}{A}+\frac{D^{''}}{D}-\frac{A^{'}}{A}(\frac{B^{'}}{B}-\frac{D^{'}}{D})-\frac{D^{'}}{D}\frac{B^{'}}{B}]=8\pi(p_{t}C^2-2\eta\sigma_{22})+4\frac{s^{2}}{D^{2}}
\end{equation}

\begin{equation}
-\frac{D^{2}}{A^{2}}[\frac{\ddot{B}}{B}+\frac{\ddot{C}}{C}-\frac{\dot{A}}{A}(\frac{\dot{B}}{B}+\frac{\dot{C}}{C})+\frac{\dot{B}\dot{C}}{BC}]+\frac{D^{2}}{B^{2}}[\frac{A^{''}}{A}+\frac{C^{''}}{C}-\frac{A^{'}}{A}(\frac{B^{'}}{B}-\frac{C^{'}}{C})-\frac{C^{'}}{C}\frac{B^{'}}{B}]=8\pi(p_{t}D^{2}-2\eta\sigma_{33})+4\frac{s^{2}}{C^{2}}
\end{equation}

and
\begin{equation}
\frac{1}{AB}(\frac{\dot{C^{'}}}{C}+\frac{\dot{D^{'}}}{D}-\frac{C^{'}}{C}\frac{\dot{B}}{B}-\frac{\dot{B}}{B}\frac{D^{'}}{D}-\frac{A^{'}}{A}\frac{\dot{C}}{C}-\frac{A^{'}}{A}\frac{\dot{D}}{D})=8\pi q
\end{equation}

The gravitational energy per specific length in cylindrically symmetric space-time is defined as [28-30]
\begin{equation}
E=\frac{(1-l^{-2}\nabla^{a}r \nabla_{a}r)}{8}
\end{equation}
In principle, $E$ is the charge associated with a general current which combines the energy-momentum of the matter and gravitational waves. It is usually referred in the literature as $C$-energy for the cylindrical symmetric space-time. For cylindrically symmetric model with killing vectors the circumference radius $\rho$ and specific length $l$ are defined as [28-30]\\

$\rho^{2}=\xi_{(1)a}\xi_{(1)}^{a}$~~~,~~~$l^{2}=\xi_{(2)a}\xi_{(2)}^{a}$,~~~~~
so that $r=\rho l$ is termed as areal radius.\\

For the present model with the contribution of electromagnetic field in the interior region the $C-$energy takes the form

\begin{equation}
E'=\frac{l}{8}+\frac{1}{8D}[\frac{1}{A^{2}}(C\dot{D}+\dot{C}D)^{2}-\frac{1}{B^{2}}(CD^{'}+C^{'}D)^{2}]+\frac{s^{2}}{2C}
\end{equation}\\

It should be noted that the above energy is also very similar to Tabu's mass function in the plane symmetric space-time[31]\\
The exterior space-time manifold $(M^{+})$ of the cylindrical surface  $\Sigma$ is described by the metric in the retarded time co-ordinate as [32,33]
		
\begin{equation}
ds_{+}^{2}=-(\frac{-2M(v)}{R}+\frac{Q^{2}(v)}{R^{2}})dv^{2}-2dRdv+R^{2}(d\phi^{2}+\lambda^{2}dz^{2})
\end{equation}
where $v$ is the usual retarded time, $M(v)$ is the total mass inside $\Sigma$, $Q(v)$ is the total charge bounded by $\Sigma$ and $\lambda$ is an arbitrary constant. Further, from the point of view of the interior manifold $(M^{-})$ the bounding three surface $\Sigma$ (comoving surface) is described as 
\begin{equation}
f_{-}(t,r)=r-r_{\Sigma}=0
\end{equation}
and hence the interior metric on $\Sigma$ takes the form
\begin{equation}
ds_-^2\stackrel{\Sigma}{=}-d\tau^2 +C^2dz^2+D^2d\phi^2
\end{equation}
where 
\begin{equation}
d\tau\stackrel{\Sigma}{=}Adt,
\end{equation}\\

defines the time co-ordinate on $\Sigma$ and $\stackrel{\Sigma}{=}$ by notation implies the equality of both sides on the surface $\Sigma$.\\
Similarly, from the perspective of the exterior manifold the boundary three surface $\Sigma$ is characterized by 

\begin{equation}
f_{+}(v,R)\equiv R-R_{\Sigma}(v)=0
\end{equation}

so that the exterior metric on $\Sigma$ takes the form

\begin{equation}
ds_{+}^{2}\stackrel{\Sigma}{=}-(\frac{-2M(v)}{R}+\frac{Q^{2}(v)}{R^{2}}+\frac{2dR_{\Sigma}(v)}{dv})dv^{2}+R^{2}(d\phi^{2}+\lambda^{2}dz^{2})
\end{equation}
Here by notation we write $[x^{+\mu}]=[v,R,\phi, z]$

\section{Junction conditions}

In order to have a smooth matching of the interior and exterior manifolds over the bounding three surface (not a surface layer), the following conditions due to Darmois [11]
are to be satisfied:\\
(i)The continuity of the first fundamental form i.e.

\begin{equation}
(ds^2)_{\Sigma}=(ds^2_{-})_{\Sigma}=(ds^{2}_{+})_{\Sigma}
\end{equation}

(ii)The continuity of the second fundamental form i.e $K_{ij}d\xi^{i}d\xi^{j}$. This implies the continuity of the extrinsic curvature $K_{ij}$ over the hypersurface [11] i.e.  
 \begin{equation}
[K_{ij}]\equiv K_{ij}^+ -K_{ij}^-=0
\end{equation}

where $K_{ij}^\pm$ is given by

\begin{equation}
K_{ij}^\pm=-n_\sigma^\pm[\frac{\partial^2x_\pm ^\sigma}{\partial\xi^i \partial\xi^j}+\Gamma_{\mu\nu}^\sigma \frac{\partial x_\pm^\mu}{\partial\xi^i}\frac{\partial x_\pm^\nu}{\partial\xi^j}],~~~(\sigma,~\mu,~\nu~=0,1,2,3)
\end{equation}

In the above expression for extrinsic curvature, $n_{\sigma}^{\pm}$ are the components of the outward unit normal to the hyper-surface with respect to the manifolds $M^{\pm}$ (i.e. in the co-ordinates   $x^{\pm \mu}$) and have explicit expressions \\
$n_\sigma^- \stackrel{\Sigma}{=}(0,B,0,0)~~and~~~~n_\sigma^+ \stackrel{\Sigma}{=}\mu(\frac{-dR}{dv},1,0,0)$
with $\mu=[\frac{-2M(v)}{R}+\frac{Q^{2}(v)}{R^{2}}+2\frac{dR}{dv}]^{\frac{-1}{2}}$\\

Also in the above the christoffel symbols are evaluated for the metric in $M^{-}$ or $M^{+}$ accordingly and we choose $\xi^{0}=\tau$,~~~$\xi^{2}=z$,~~~$\xi^{3}=\phi$ as the intrinsic co-ordinates on   $\Sigma$ for convenience.\\                              

The  continuity of the 1st fundamental form  gives 

\begin{equation}
C(t,r_{\Sigma})\stackrel{\Sigma}{=}R_{\Sigma}(v)~~~,~~~D(t,r_{\Sigma})\stackrel{\Sigma}{=}\lambda R_{\Sigma}(v)
\end{equation}

\begin{equation}
\frac{dt}{d\tau}=1/A~~~~\frac{dv}{d\tau}=\mu 
\end{equation}

Now the  non vanishing components of extrinsic curvature $K^{\pm}_{ij}$ are\\
\begin{equation}
 K_{00}^-{=}-(\frac{A^\prime}{AB})_{\Sigma}
\end{equation}
\begin{equation}
K_{00}^+{=}[(\frac{d^{2}v}{d\tau^{2}})(\frac{dv}{d\tau})^{-1}-(\frac{dv}{d\tau})(\frac{M}{R^{2}}-\frac{Q^{2}}{R^{3}})]_{\Sigma}
\end{equation}\\
 \begin{equation}
K_{22}^-{=}(\frac{C{C}^\prime}{B})_{\Sigma}
\end{equation}
\begin{equation}
K_{33}^-{=}(\frac{D{D}^\prime}{B})_{\Sigma}
\end{equation}

\begin{equation}
K_{22}^+{=}[R(\frac{dR}{d\tau})-\frac{(dv)}{d\tau}(2M-\frac{Q^{2}}{R})]_{\Sigma}=\lambda ^{-2}K^{+}_{33}
\end{equation}\\
and\\

Hence continuity of the extrinsic curvature together with equations (32) and (33) gives the following relations over $\Sigma$ [33,34] 
\begin{equation}
M(v)\stackrel{\Sigma}{=}\frac{R}{2}[(\frac{\dot{R}}{A})^{2}-(\frac{R^{'}}{B})^{2}]+\frac{Q^{2}}{2R}
\end{equation}\\
\begin{equation}
E\stackrel{\Sigma}{=}\frac{l}{8}+\lambda M
\end{equation}\\
and 
\begin{equation}
q\stackrel{\Sigma}{=}p_{r}-\frac{2\eta \sigma_{11}}{B^{2}}-\frac{s^{2}}{2\pi c^{4}}(\frac{1}{\lambda^{2}}-1/4
)
\end{equation}\\

Thus equations (39) gives the total mass inside the boundary surface $\Sigma$ while equation (40) shows the linear relationship between the $C$ energy for the cylindrically symmetric space-time with the bounding mass over $\Sigma$. Further, equation (41) shows a linear relationship among the fluid parameters $(p_{r}, \eta, q)$ on the bounding surface $\Sigma$. Hence radial pressure is in general non zero on the bounding surface due to dissipative nature of the fluid and the charge on the bounding surface. But when dissipative components of the fluid are switch off then the above result (uncharged) agrees with the results of Herrera etal [8]. Also it should be noted that the radial pressure on the boundary does not depend on the charge bounded by $\Sigma$, it depends only on the charge on the surface $\Sigma$.\\

\section{Analysis of Dynamical equations:}
From the conservation of energy-momentum i.e. $(T^{\alpha\beta}+E^{\alpha\beta})_{;\beta}=0$ we can have two zero scalars namely
$(T^{\alpha\beta}+E^{\alpha\beta});_{\beta}v_{\alpha}~~~and~~~(T^{\alpha\beta}+E^{\alpha\beta});_{\beta}\chi_{\alpha}$\\
Using equations (2) and(8) the explicit expressions for these two scalars are
\begin{equation}
\frac{\dot{\rho}}{A}+\frac{\dot{B}}{A}(\frac{\rho}{B}+\frac{p_{r}}{B}-2\eta\sigma^{11})+\frac{\dot{C}}{A}(\frac{p_{\bot}}{C}+\frac{\rho}{C}-2\eta\sigma^{22})+\frac{\dot{D}}{A}(\frac{\rho}{D}+\frac{p_{\bot}}{D}-2\eta\sigma^{33})+\frac{q^{'}}{B}+\frac{q}{B}(2\frac{A^{'}}{A}+\frac{C^{'}}{C}+\frac{D^{'}}{D})=0
\end{equation}
and
\begin{eqnarray}
(\frac{p_{r}}{B^{2}}-2\eta\sigma^{11})^{'}+\frac{\dot{q}}{AB}+\frac{q}{AB}(\frac{\dot{C}}{C}+\frac{\dot{D}}{D})+\frac{A^{'}}{A}(\frac{\rho}{B^{2}}+\frac{p_{r}}{B^{2}}-2\eta\sigma^{11})+\frac{B^{'}}{B}(\frac{p_{r}}{B^{2}}-2\eta\sigma^{11})+\frac{C^{'}}{C}(\frac{p_{r}}{B^{2}}
\nonumber
\\
-\frac{p_{\bot}}{B^{2}}-2\eta\sigma^{11}-2\eta\sigma^{22}\frac{C^{2}}{B^{2}})+\frac{D^{'}}{D}(\frac{p_{r}}{B^{2}}-\frac{p_{\bot}}{B^{2}}-2\eta\sigma^{11}+2\eta\sigma^{33}\frac{D^{2}}{B^{2}})-\frac{ss^{'}}{\pi C^{2}D^{2}B}=0 
\end{eqnarray}\\

Now following the formulation of Misner and Sharp [2], we introduce the proper time derivative and proper radial derivative as 
\begin{equation}
D_{T}=\frac{1}{A}\frac{\delta}{\delta t}~~~~~~and~~~~ D_{R}=\frac{1}{R^{'}}\frac{\delta}{\delta r}
\end{equation}

so that the fluid velocity in the collapsing situation, can be defined as [35]
\begin{equation}
U=D_{T}(R)=D_{T}(C) <0
~~~and~~~ V=D_{T}(Rr)=D_{T}(D) <0
\end{equation}
Using equations (17)-(22) and(45), we can obtain the acceleration of a collapsing matter inside $\Sigma$  as 

\begin{equation}
D_{T}(U)=-4\pi R(p_{r}-\frac{4\eta \sigma}{\sqrt{3}})+\tilde{E}\frac{A^{'}}{AB}+\frac{s^{2}}{R^{3}}(2+\frac{1}{2\lambda})-\frac{1}{R^{2}\lambda}(E^{'}-\frac{l}{8})
\end{equation}
Now combining (43) and (46) we obtain
\begin{eqnarray}
(\rho+p_{r}-\frac{4\eta\sigma}{\sqrt{3}})D_{T}(U)&=&(\rho+p_{r}-\frac{4\eta\sigma}{\sqrt{3}})[\frac{1}{R^{2}\lambda}(E^{'}-\frac{l}{8})+4\pi R(p_{r}-\frac{4\eta \sigma}{\sqrt{3}})-\frac{s^{2}}{R^{3}}(2+\frac{1}{2\lambda})]-\tilde{E}^{2}[D_{R}(p_{r}
\nonumber
\\
&-&\frac{4\eta\sigma}{\sqrt{3}})+\frac{2}{R}(p_{r}-p_{\bot}-2\sqrt{3}\eta\sigma)-\frac{s}{\pi R^{4}}D_{R}(s)]-\frac{2q\tilde{E}}{A}(\frac{\dot{B}}{B}+\frac{\dot{C}}{C})-\frac{\dot{q}\tilde{E}}{A}
\end{eqnarray}\\

Using (22) and the junction condition $D\stackrel{\Sigma}{=}\lambda C$,  we write [24]
\begin{equation}
\tilde{E} =\frac{C^{'}}{B}=[U^{2}+\frac{s^{2}}{\lambda c^{2}}-\frac{2}{\lambda c}(E^{'}-\frac{l}{8})]^{\frac{1}{2}}
\end{equation}

Hence using the field equations for the interior manifold we obtain the time rate of change of C-energy as 
\begin{equation}
D_{T}E^{'}=-4\pi R^{2}\lambda[(p_{r}-\frac{4\eta\sigma}{\sqrt{3}})U +q\tilde{E}]+\frac{s^{2}\dot{C}}{R^{2}A}(2\lambda-\frac{1}{2})
\end{equation}
Also the above equation can be interpreted as the variation of the total energy inside the  collapsing cylinder. Note that due to negativity, of the fluid velocity $v$ the first term on the r.h.s will contribute to the energy of the system provided the radial pressure is restricted as $p_{r}>\frac{4\eta\sigma}{\sqrt{3}}$. Due to negativity, the second term indicates an outflow of energy in the form of radiation during the collapsing process. The third term is coulomb-like force term and it will increase the energy of the system provided $\lambda~>~\frac{1}{4}$.\\
Further, using the Einstein field equations (16), (20) and the expression for C- energy in equation (22), the radial derivative of the C energy takes the form
\begin{equation}
D_{R}E'=4\pi\rho R^{2}\lambda+\frac{s^{2}}{R^{2}}(2\lambda-\frac{1}{2})+\frac{s}{R}D_{R}(s)+\frac{4\pi q BR^{2}\lambda}{R'}D_{T}(C)+\frac{1}{8\rho R'}
\end{equation}

This radial derivative can be interpreted as the energy variation between the adjacent cylindrical surfaces within the matter distribution. The first term on the r.h.s. is the usual energy density of the fluid element while the second term and third terms are the conditions due to the electromagnetic field. The fourth term represents contribution due to the dissipative heat flux and the last term will increase or decrease the energy of the system during the collapse of the cylinder provided $R^{'}> ~or~<~0$\\

Finally, the collapse dynamics is completely characterized by the equation of motion in equation (47). Normally, for collapsing situation $D_{T}U$ should be negative, i.e, indicating an inward radial flow of the system. Consequently, terms on the r.h.s (of eq.(47)) contributing negatively favours the collapse and positive terms oppose the collapsing process.  In an extreme situation the system will be in hydrostatic equilibrium if terms of both signs balance each other. Further, from dimensional analysis the factor $(\rho+p_{r}-\frac{4\eta\sigma}{\sqrt{3}})$ can be considered as an inertial mass density, independent of heat flux contribution. The first term on the r.h.s. of eq. (47) can be identified as the gravitational force, indicating the effects of specific length and electric charge in the gravitational contribution. The second term has three contributing components- the pressure gradient (which is negative), local anisotropy of the fluid and electromagnetic field term. The remaining terms represent the heat flux contribution and due to negativity they seem to leave the system along the radial outward streamlines.\\

\section{Causal Thermodynamics: The Transport equation} 
In causal thermodynamics due to Miller-Israel-Stewart, the transport equation for heat flow is given by [21]

\begin{equation}
\tau h^{ab}V^{c}q_{b;c}+q^{a}=-\kappa h^{ab}(T_{,b}+a_{b}T)-\frac{1}{2}\kappa T^{2}(\frac{\tau V^{b}}{\kappa T^{2}})_{;b}q^{a}
\end {equation}

where $h^{ab}=g^{ab}+V^{a}V^{b}$ is the projection tensor of the $3-$surface orthogonal to the unit time-like vector $V^{a}$, $\kappa$ represents the thermal conductivity, $T$ is the temperature, $\tau$ denotes the relaxation time and $a_{b}T$ is the inertial term due to Tolman. 
Now due to cylindrical symmetry, the above transport equation (51)simplifies to  

\begin{equation}
\tau\dot{q}=-\frac{1}{2A}\kappa \frac{qT^{2}}{\tau}(\frac{\tau}{\kappa T^{2}})-q[\frac{3U}{2R}+G+\frac{1}{\tau}]-\frac{\kappa \tilde{E}D_{R}T}{\tau}-\frac{\kappa TD_{T}U}{\tau\tilde{E}}-\frac{\kappa T}{\tau\tilde{E}R^{2}}[\frac{1}{\lambda}(E^{'}-\frac{l}{8})+4\pi R^{3}(p_{r}-\frac{4\eta\sigma}{\sqrt{3}})-\frac{S^{2}}{R}(2+\frac{1}{2\lambda})]
\end {equation}
with $G=\frac{1}{A}(\frac{\dot{B}}{B}-\frac{\dot{C}}{C})$\\ 

Now considering proper derivatives in equation(44) of the above equation and using the field velocity (in eq.(45)), and equation of motion (i.e. eq. (47)) one obtains the effects of heat flux or dissipation in the collapsing process as \\

\begin{eqnarray}
(1-\alpha)(\rho+ p_{r}-\frac{4\eta\sigma}{\sqrt{3}})D_{T}U =(1-\alpha)F_{grav}+F_{hyd}+ \alpha \tilde{E}^{2}[D_{R}p_{r}+2(p_{r}-p_{\bot}-2\sqrt{3}\eta\sigma)\frac{1}{R}
\nonumber 
\\
-\frac{SD_{R}(S)}{\pi R^{4}\lambda^{2}}]-\tilde{E}[D_{T}q+2qG+\frac{4qU}{R}]+\alpha \tilde{E}[D_{T}q+\frac{4qU}{R}+2qG]
\end{eqnarray}\\

with 
\begin{equation}
\alpha=\frac{\kappa T}{\tau}(\rho+p_{r}-\frac{4\eta\sigma}{\sqrt{3}})^{-1}
\end{equation}

\begin{equation}
F_{grav}=-(\rho+p_{r}-\frac{4\eta\sigma}{\sqrt{3}})[(E^{'}-\frac{l}{8})\frac{1}{\lambda}+4\pi p_{r}R^{3}-(2+\frac{1}{2\lambda})\frac{S^{2}}{R}](\frac{1}{R^{2}})
\end{equation}

\begin{equation}
F_{hyd}=\tilde{E}^{2}[D_{R}(p_{r}-\frac{4\eta\sigma}{\sqrt{3}})+\frac{2}{R}(p_{r}-p_{\bot}-2\sqrt{3}\eta\sigma)-\frac{S}{\pi R^{4}}D_{R}(S)]
\end{equation}

The l.h.s. of equation (53) can be interpreted as Newtonian force $F$ with $(\rho+p_{r})(1-\alpha)$ as the inertial mass density. So as $\alpha\rightarrow 1$, $F\rightarrow 0$ i.e. there is no inertial force and collapse will be inevitable due to gravitational attraction. Further, the inertial mass density decreases as long as $0~<~\alpha~<1$  and it increases for $\alpha~>~1$. Moreover, due to equivalence principle the gravitational mass also decrease or increase according as $\alpha~<~or~>~1$ and gives a clear distinction between the expanding and collapsing process due to dynamics of dissipative system. Note that although the gravitational force is affected by the same factor $(1-\alpha)$ but the hydrodynamical force is free from it. Further, combination of all these terms on the r.h.s of equation (53) results the l.h.s i.e. $(1-\alpha)(\rho+p_{r}-\frac{4\pi\sigma}{\sqrt{3}})D_{T}U~<0$, there will be gravitational collapse while there will be expansion if the l.h.s to be positive. Interestingly, if $\alpha$ continuously decreases from a value larger than unity to one less than unity, then there will be a phase transition (collapse to expansion) and bounce will occur. As a result, there will be loss of energy of the system and collapsing cylinder with non-adiabatic source causes emission of gravitational radiation. Therefore, there will be radiation outside the collapsing cylinder and hence the choice of the exterior metric (in eq.(23)) is justified.\\


\begin{thebibliography}{}
\bibitem{Oppenheimer} J.R. Oppenheimer and H. Snyder , {\it Phys.Rev.} {\bf 56,} (1939)455 ; 
\bibitem{Misner} C.W. Misner and D. Sharp, {\it Phys.Rev.} {\bf 136} (1964) B571;
\bibitem{Vaidya} P.C.Vaidya,  {\it Proc.Indian Acad.Sci.A} {\bf 33} (1951) 264;
\bibitem{Joshi} P.S. Joshi and T.P. Singh, {\it Phys.Rev.D} {\bf 51} (1995) 6778; 
\bibitem{Debnath} U. Debnath, S. Nath and S. Chakraborty {\it Gen. Relt. Grav.} {\bf 37} (2005) 215.
\bibitem{Levi-Civita} T. Levi-Civita, {\it Rend. Accad. Lincei} {\bf 28} (1919) 101.
\bibitem{Herrera6} L. Herrara and N.O. Santos, {\it Class. Quant. Grav} {\bf 22} (2005) 2407.
\bibitem{Herrera7} L. Herrera, N.O. Santos and M.A.H. Maccallum, {\it Class. Quant. Grav} {\bf 24} (2007) 1033.
\bibitem{Di Prisco} A. Di Prisco, L. Herrera, M. Maccallum and N.O. Santos, {\it Phys. Rev. D.} {\bf 80} (2009) 064031.
\bibitem{Herrera8} L. Herrera, A. Di Prisco, J. Ospino,  {\it Gen.Relt. Grav.} {\bf 44} (2012) 2645.
\bibitem{Darmois} G. Darmois, Memorial des Sciences Mathematiques (Gautheir-Viuars, Paris,) Fasc. 25.
\bibitem{Sharif1} M. Sharif and Z. Ahmed {\it Mod.Phys.Lett.A} {\bf 22} (2007) 1493
\bibitem{Sharif2} M. Sharif and Z. Ahmed {\it Mod.Phys.Lett.A} {\bf 22} (2007) 2947.
\bibitem{Sharif3} M. Sharif and Z. Ahmed {\it Gen.Relt. Grav.} {\bf 39} (2007) 1331.
\bibitem{Herrera3} L. Herrera , G. Le Denmat, G. Marcilhacy,  N.O. Santos {\it Int. J. Mod. Phys.D} {\bf 14} (2005) 657.
\bibitem{Kurita} Y. Kurita and K. Nakao, {\it Phys.Rev. D} {\bf 73} (2006) 064022.
\bibitem{Herrera1} L. Herrera, A. DiPrisco, J.Martin, J.Ospino, N.O. Santos and O.Troconis  {\it Phys.Rev.D} {\bf 69} (2004) 084026.
\bibitem{Herrera2} L. Herrera, N.O. Santos,  {\it Phys. Rev. D.} {\bf 70} (2004) 084004.
\bibitem{Mitra} A. Mitra, {\it Phys.Rev.D.} {\bf 74} (2006) 024010.
\bibitem{Chan} R. Chan, {\it Astron.Astrophys.} {\bf 368} (2001) 325.
\bibitem{Herrera4} L. Herrera,  {\it Int. J. Mod Phys.D} {\bf 15} (2006) 2197.
\bibitem{Herrera5} L. Herrera, A. Di Prisco, E.Fuenmayor and O. Troconics, {\it Int. J. Mod. Phys. D} {\bf 18} (2009) 129.
\bibitem{Sharif8} M. Sharif and Z. Rehmat, {\it Gen.Relt.Grav.} {\bf 42} (2010) 1795.
\bibitem{Sharif} M. Sharif and G. Abbas, {\it Astrophys.Space Sci} {\bf 335} (2011) 515.
\bibitem{Chakraborty} S. Chakraborty and S. Chakraborty  {\it Gen.Relt. Grav.} {\bf 46}, (2014) 1784; {\it Annals of Physics} {\bf 364}, (2016) 110
\bibitem{Sharif} M. Sharif and G. Abbas, {\it J.Phys.Soc.Jpn} {\bf 80} (2011) 104002.
\bibitem{Prisco} A.Di Prisco etal, {\it Phys.Rev.D} {\bf 76} (2007) 064017.
\bibitem{Throne} K.S Throne, {\it Phys.Rev.} {\bf 138} (1965) B251.
\bibitem{Chiba} T. Chiba, {\it Prog.Theo.Phys.} {\bf 95} (1996) 321.
\bibitem{Hayward} S.A. Hayward, {\it Class.Quant.Grav.} {\bf 17} (2000) 1749.
\bibitem{Sharif5} M. Sharif and Z. Rehamat {\it Gen.Relt.Grav} {\bf 42} (2010) 1795.
\bibitem{Guang} H. Chao-Guang  {\it Acta Phys.Sin.} {\bf 4} (1995) 617.
\bibitem{Sharif6} M. Sharif and M. Azam {\it Jour.of cosmology and Astro.Physics} {\bf 02} (2012) 043
\bibitem{Guha}S.Guha and R.Banerji {\it Int J Theorn Phys.} {\bf 53} (2014)2332
\bibitem{Sharif7} M. Sharif and S. Fatima {\it Gen.Relt.Grav.} {\bf 43} (2011) 127























































































































































































































































































































































































































































































































































































































































































































































































































































































































































































































































































































































































































































































































\end{thebibliography}
\end{document}